\documentclass[letterpaper,twocolumn]{IEEEtran}

\usepackage{graphicx}
\usepackage[cmex10]{amsmath}
\usepackage{amsfonts}
\usepackage{latexsym}
\usepackage{amssymb}
\usepackage{cite}
\usepackage{slashbox}
\usepackage{multirow}
\usepackage{algorithmic,algorithm}
\usepackage{hyperref}
\usepackage{mathtools}
\usepackage{indentfirst}
\usepackage{bm}

\newtheorem{theorem}{Theorem}
\newtheorem{lemma}{Lemma}
\newtheorem{proposition}{Proposition}

\ifCLASSINFOpdf
\else
\fi
\begin{document}
%
\title{Self-Interference Suppression for the Full-Duplex Wireless Communication with Large-Scale Antenna}


\author{\IEEEauthorblockN{Pengbo Xing\IEEEauthorrefmark{0},
Ju Liu, \emph{Senior Member}, \emph{IEEE}\IEEEauthorrefmark{0},
Chao Zhai\IEEEauthorrefmark{0},
Xinhua Wang\IEEEauthorrefmark{0}, and
Lina Zheng\IEEEauthorrefmark{0}}

\IEEEauthorblockA{\IEEEauthorrefmark{0}
\thanks{This work was supported by the National Natural Science Foundation of China (61371188), the Research Fund for the Doctoral Program of Higher Education (20130131110029), the Open Research Fund of the National Mobile Communications Research Laboratory (2012D10), the China Postdoctoral Science Foundation (2014M560553), and the Special Funds for Postdoctoral Innovative Projects of Shandong Province (201401013).}
\thanks{The authors are with the School of Information Science and Engineering, Shandong University, Jinan, China. Pengbo Xing and Ju liu are also with the National Mobile Communications Research Lab., Southeast University, Nanjing, China (e-mail: juliu@sdu.edu.cn).
}}}

%



\IEEEtitleabstractindextext{%
\begin{abstract}
In this letter, we proposed a shared-antenna full-duplex framework for the multiuser MIMO system with a large-scale antenna array exploited at the base station to transmit and receive signals simultaneously. It has the merits of both the full-duplex system and the time-division duplex massive MIMO system, i.e., the high spectral efficiency and the channel reciprocity. We focus on the zero-forcing (ZF) and the maximal-ratio transmission/maximal-ratio combining (MRT/MRC) linear processing methods, which are commonly used in the massive MIMO system. As the main finding, we prove that the self-interference in the completely dependent uplink and downlink channels of this full-duplex system can be suppressed by the large-scale antenna array.
\end{abstract}

}

\maketitle

\IEEEdisplaynontitleabstractindextext

%
\IEEEpeerreviewmaketitle

\section{Introduction}
%
%
%
%
\IEEEPARstart{L}{arge-scale} MIMO (massive MIMO) and in-band full-duplex have been attracting lots of interests from academia and industry nowdays \cite{Larsson,Sabha,Lulu,Rusek,Riihonen,Everett,Duarte,Ngo1,Ngo2,Hong,Yang,Hoydis}, because both of them can achieve high spectral efficiency. In a traditional full-duplex system \cite{Riihonen,Everett,Duarte}, a separate antenna or antenna array is employed to receive the signals from the remote transmitters, meantime the self-interference (SI) is also received from its own transmitter nearby. Compared with the desired signals, the SI is much stronger, thereby it must be suppressed before decoding the signals from the remote transmitters. Many studies have shown that the SI can be mitigated by active or passive suppressions, in the wireless propagation, analog circuit, or digital domain \cite{Sabha}, \cite{Riihonen,Everett,Duarte}. Meanwhile, a massive MIMO system usually works in the half-duplex mode, such as the time-division duplex (TDD) and employs a single antenna array so that the channel reciprocity can be assumed \cite{Lulu, Rusek, Ngo1}. In a TDD massive MIMO system, the downlink channel matrix is a transpose of the uplink channel matrix, therefore, the channel reciprocity can help reduce the burden of channel state information (CSI) feedback. With a large-scale antenna processing method, the interference and noise will be vanished when the scale of antenna array is large enough \cite{Ngo1}. The combination of the full-duplex and the massive MIMO is actually becoming a hot topic recently \cite{Ngo2}.

However, this combination faces the conflicts between the separate antenna arrays in a full-duplex system and the single antenna array in a massive MIMO system. 
Two types of full-duplex antenna interfacing methods have been summarized in \cite{Sabha}, namely, the \emph{separate-antenna} and the \emph{shared-antenna} interfacing. The shared-antenna interfacing can reduce the antenna pair to a single antenna with a circulator, which performs the directional routing for the electromagnetic wave in a full-duplex wireless transceiver. The circulator has been also introduced in \cite{Hong}. 
In this letter, we propose a framework for the shared-antenna full-duplex massive multiuser MIMO system, it has a single antenna array equipped with circulators. This model has the merits of both the full-duplex and the TDD massive MIMO, i.e., the high spectral efficiency and the channel reciprocity. When the antenna array scale $M$ is large enough, we find that the SI in a shared-antenna full-duplex massive MIMO system can be suppressed by the linear processing methods regardless of the complete dependence of the uplink and the downlink channels. The simulation results are provided to verify the effects of our proposed model in suppressing the SI.

 \emph{Notation:}
The superscripts $[\cdot]{^*}$, $[\cdot]^T$, and $[\cdot]^H$ stand for the conjugate, transpose, and Hermitian transpose of matrices or vectors, respectively. The $tr(\cdot)$ represents the trace of a square matrix. The expectation operator is denoted as $\mathbb{E}\{\cdot\}$. The $[\cdot]_{ij}$ represents the $(i,j)$th entry of a matrix. The $\xrightarrow{a.s.}$ represents \emph{almost sure convergence}, and $u_n=O(w_n)$ denotes that there exists a constant $P$, such that $u_n \leq Pw_n$ for all $n$.
\section{System Model}

\subsection{Shared-Antenna Full-Duplex Massive MU-MIMO Model}

Fig. 1 shows the shared-antenna full-duplex massive MU-MIMO system, where $K$ user equipments (UEs) are served by a base station (BS). Each UE is equipped with one antenna, while the BS has $M$ antennas. Both the UEs and the BS are working in the full-duplex mode, i.e., they can transmit and receive signals using the same time-frequency resource. Like the separate-antenna full-duplex system \cite{Everett}, the physical isolation is exploited between the BS antennas to avoid the front end saturation at the receiver. Therefore, the $M\times1$ uplink received signal vector at the BS is given as
\begin{equation}\label{Eqn:yBS}
{\bm y_{\rm BS}} = \sqrt {p_{\rm u}} \bm G{\bm x_{\rm u}} + \sqrt {p_{\rm d}} {\bm G_{\rm s}}\bm s + {\bm n},
\end{equation}
where $\bm G$ is the $M \times K$ channel matrix between the $K$ users and the BS; ${\bm x_{\rm u}}$ is the vector of symbols transmitted by the $K$ users, and it has unit variance $\mathbb{E}\{{\bm x_{\rm u}\bm x{_{\rm u}^H}}\}=\bm I_{K}$; The normalized uplink power is $p_{\rm u}$; $\bm G_{\rm s}$ is the $M \times M$ SI channel matrix between the BS transceivers;
${\bm{s}}$ is the downlink precoded vector transmitted to the $K$ users, and it has a total expected power $\mathbb{E}\{{\bm s^H\bm s}\}=1$; The normalized downlink power is $p_{\rm d}$; $\bm n$ is an additive complex Gaussian noise vector with zero mean and unit variance.

The channel matrix $\bm G$ incorporates the small-scale Rayleigh fading and the large-scale fading, so we have
{\setlength\abovedisplayskip{5pt}
\setlength\belowdisplayskip{5pt}
\begin{align}\label{Eqn:GHD}
\bm G=\bm H \bm D^{1/2},
\end{align}}
where $\bm H$ is the $M \times K$ matrix with i.i.d. entries, which represent the small-scale fading between the $K$ users and the BS, and the entry $h_{mk}$ is a circularly symmetric complex Gaussian (CSCG) random variable with zero mean and unit variance, i.e., $h_{mk}\sim \mathcal{CN}(0,1)$; $\bm D$ is a $K \times K$ diagonal matrix of the large-scale fading, where $[\bm D]_{kk}=\beta_k$.
\begin{figure}
\centering \scalebox{0.50}{\includegraphics[width=6.5in]{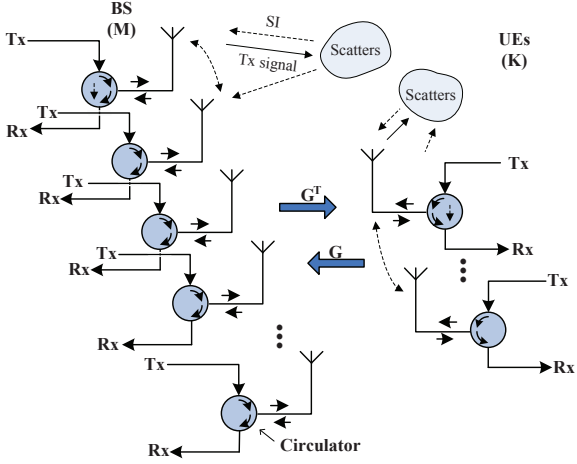}}
\caption{ The shared-antenna full-duplex massive MU-MIMO model.} \label{fig1}
\end{figure}

The SI channel matrix $\bm G_{\rm s}$ models the direct-path interference and the reflected-path interference factors. Unlike the single shared-antenna, for the shared-antenna array, the direct-path SI comes from not only the circulator leakage, but also the antennas inter-coupling if they are not ideally isolated; The reflected-path SI is from not only its own reflection, but also the others. We assume that the direct-path SI is deterministic and the reflected-path SI is random. Therefore, the SI channel matrix can be modeled as
\begin{align}
\bm G_{\rm s}&={\bar {\bm G}}_{\rm s}+{\tilde {\bm G}}_{\rm s},
\end{align}
where ${\bar {\bm G}}_{\rm s}$ is an $M \times M$ complex deterministic matrix with $[{\bar {\bm G}}_{\rm s}]_{ij}=c_{ij}$, which represents the antennas intra-leakage and inter-coupling coefficient (direct-path). ${\tilde {\bm G}}_{\rm s}$ is an $M \times M$ complex random matrix (reflected-path) \cite{Riihonen},\cite{Ngo2}. Like the signal channel matrix $\bm G$, ${\tilde {\bm G}}_{\rm s}$ can be expressed as
\begin{align}
{\tilde {\bm G}}_{\rm s}={\tilde {\bm H}}_{\rm s} {\tilde {\bm D}}{_{\rm s}^{1/2}},
\end{align}
where ${\tilde {\bm H}}_{\rm s}$ is an $M \times M$ matrix with i.i.d. entries, which represents the SI small-scale fading between the BS transceivers, it is caused by the reflection from the scatterers. We suppose $[{\tilde {\bm H}}_{\rm s}]_{ij}={\tilde h}_{ij}$, where ${\tilde h}_{ij}$ is a CSCG random variable with zero mean and unit variance; ${\tilde {\bm D}}_{\rm s}$ is an $M \times M$ diagonal matrix of the large-scale fading coefficients, where $[{\tilde {\bm D}}_{\rm s}]_{mm}={\beta}$; We assume that all the SI large-scale fading coefficients have the same value ${\beta}$, because the size of antenna array is much smaller than the distance of the scatterers from the array.

\subsection{Uplink Decoding Under the Downlink Precoding SI}
As mentioned above, the downlink precoding can be used in the data transmission phase so that the strength of radio signal at the corresponding UE can be enhanced. However, the CSI should be available at the transmitter before the precoding. In this letter, we suppose the BS knows the CSI perfectly through a training process. The BS transmitters are assumed to adopt the ZF or the MRT precoding scheme for the downlink transmission. Correspondingly, the BS receivers use the ZF or the MRC linear processing method for the uplink data reception.

Suppose $\bm x_{\rm d}$ is the $K\times 1$ data vector sent to the users by the BS, and $\bm s =\bm A\bm x_{\rm d}$, where $\bm A$ is the precoding matrix \cite{Ngo2},
{\setlength\abovedisplayskip{5pt}
\setlength\belowdisplayskip{5pt}
\begin{eqnarray}\label{Eqn:A}
\bm A=
\left\{
\begin{aligned}
&\alpha_{\rm ZF}{{{\bm G}}^*}({\bm G}^T{\bm G}^*)^{-1},&&\quad\mbox{for ZF},\\
&\alpha_{\rm MRT}{{{\bm G}}^*},&&\quad \mbox{for MRT,}
\end{aligned}
\right.
\end{eqnarray}}
and $\mathbb{E}\{{\bm x_{\rm d}\bm x{_{\rm d}^H}}\}=\bm I_{K}$; The normalization factors $\alpha_{\rm ZF}$ and $\alpha_{\rm MRT}$ satisfy $\mathbb{E}\{{\bm s^H\bm s}\}=1$, so we have $\alpha_{\rm ZF} \!\triangleq \!\sqrt {M-K\over {\sum_{k=1}^K{\beta}{_k^{-1}}}}$, and $\alpha_{\rm MRT} \!\triangleq \!\sqrt {1\over {M\sum_{k=1}^K{\beta}{_k}}}$ \cite{Ngo2, Yang}. Then, Eq. (\ref{Eqn:yBS}) can be rewritten as
{\setlength\abovedisplayskip{5pt}
\setlength\belowdisplayskip{5pt}
\begin{align}
{\bm y_{\rm BS}} = \sqrt {p_{\rm u}} \bm G{\bm x_{\rm u}} + \sqrt {p_{\rm d}} {\bm G_{\rm s}}\bm A \bm x_{\rm d} + {\bm n}.
\end{align}}
In order to obtain the uplink data sent from the users, the BS receiver will multiply a linear matrix $\bm W^T$ before the received signal, which is a function of the signal channel \cite{Ngo2}. The linear receiver matrix can be given as
{\setlength\abovedisplayskip{5pt}
\setlength\belowdisplayskip{5pt}
\begin{eqnarray}\label{Eqn:WT}
\bm W^T=
\left\{
\begin{aligned}
&({{\bm G}}^H {{\bm G}})^{-1}{{\bm G}}^H,&&\quad\mbox{for ZF},\\
&{{\bm G}}^H, &&\quad\mbox{for MRC}.
\end{aligned}
\right.
\end{eqnarray}}
With the linear receiver at the BS, the uplink signal transmitted from the UEs can be decoded from the received signal ${\bm y_{\rm BS}}$. The uplink decoded signal can be expressed as
\begin{align}\label{Eqn:r}
\bm r&=\bm W^T{\bm y_{\rm BS}}\nonumber\\
&=\sqrt {p_{\rm u}} \bm W^T\bm G{\bm x_{\rm u}} + \sqrt {p_{\rm d}}\bm W^T {\bm G_{\rm s}}\bm A \bm x_{\rm d} + \bm W^T{\bm n}.
\end{align}
Correspondingly, the uplink decoded signal for the $k$th user, namely, the $k$th element of $\bm r$, can be expressed as
\begin{align}\label{eqn:rk}
r_k=\sqrt{p_{\rm u}}\bm w{_k^T}{\bm g_k}{x_{{\rm u},k}}&+ \sqrt{p_{\rm u}}\sum_{i=1,i\neq k}^{K}{\bm w{_k^T}}{\bm g_i}{x_{{\rm u},i}}\nonumber\\
&+\sqrt{p_{\rm d}}{\bm w{_k^T}}{\bm G_{\rm s}}\bm A \bm x_{\rm d}+\bm w{_k^T}{\bm n},
\end{align}
where $\bm w{_k^T}$ and $\bm g_k$ are the $k$th columns of $\bm W^T$ and $\bm G$, respectively; $x_{{\rm u},k}$ is the $k$th element of $\bm x_{\rm u}$; $\sqrt{p_{\rm u}}\bm w{_k^T}{\bm g_k}{x_{{\rm u},k}}$ is the desired signal from the $k$th user; $\sqrt{p_{\rm u}}\sum_{i\neq k}^{K}{\bm w{_k^T}}{\bm g_i}{x_{{\rm u},i}}$ is the interference from other users; $\sqrt{p_{\rm d}}{\bm w{_k^T}}{\bm G_{\rm s}}\bm A \bm x_{\rm d}$ is the SI and $\bm w{_k^T}{\bm n}$ is the additive white noise.

\subsection{Downlink Decoding}
Like the uplink, there exists SI in the downlink of the full-duplex system. The downlink SI includes the inter-user SI and the intra-user SI. The inter-user SI occurs when the user is receiving the signal from the BS while other users nearby are transmitting signals, so it comes from the reflection and the direct coupling of the others' transmit signals. One option for the avoidance of the strong inter-user SI is to switch the duplex mode if the users are located in the same place. The corresponding UEs can switch their full-duplex mode to the half-duplex mode, which are scheduled by the BS. In this letter, we assume that all the users are located in different places and all the UEs are working in the full-duplex mode, and their inter-user SI are attenuated by the distance so that no front-end saturation happens. The intra-user SI is from the circulator leakage and its own signal reflection, and we assume that all the UEs have the intra-SI as well. Therefore, the $K \times 1$ downlink received signal vector is given as
{\setlength\abovedisplayskip{1pt}
\setlength\belowdisplayskip{5pt}
\begin{equation}\label{Eqn:yUE}
{\bm y_{\rm UE}} = \sqrt {p_{\rm d}}\bm G^T\bm A{\bm x_{\rm d}} + \sqrt {p_{\rm u}}\bm G'_{\rm s} \bm x_{\rm u} + {\bm n_{\rm d}},
\end{equation}}
where $\bm G'_{\rm s}$ is a $K \times K$ matrix, and the entry $[\bm G'_{\rm s}]_{pq}$ is $(c'_{pq}+\beta'_{pq} h'_{pq})$, where $c'_{pq}$ is the intra-leakage and inter-coupling coefficient (direct-path), $\beta'_{pq}$ is the large-scale fading of the reflected-path SI, $h'_{pq}$ is the small-scale fading, and $h'_{pq} \sim \mathcal {CN}(0,1)$; ${\bm n_{\rm d}}$ is the downlink additive complex Gaussian noise vector with zero mean and unit variance.

By substituting Eq. (\ref{Eqn:A}) into Eq. (\ref{Eqn:yUE}), and by using the asymptotic orthogonality property in \cite{Rusek}, namely, ${\bm G^T\bm G^*\over M}\xrightarrow{a.s.}\bm D$, we yield the $k$th user received signal with ZF and MRC methods
{\setlength\abovedisplayskip{3pt}
\setlength\belowdisplayskip{5pt}
\begin{equation}\label{Eqn:yk}
{y_{k}} = \sqrt {p_{\rm d}} \rho_k {x_{{\rm d},k}} + \sum_{q=1}^{K}(c'_{kq}+\beta'_{kq}h'_{kq}) x_{{\rm u},q} + {n_{{\rm d},k}},
\end{equation}}
where $\rho_k$ is the linear processing factor, it is given as
\begin{eqnarray}\label{Eqn:rhok}
\rho_k=
\left\{
\begin{aligned}
&\alpha_{\rm ZF},&&\quad\mbox{for ZF},\\
&\alpha_{\rm MRT} M\beta_k,&&\quad \mbox{for MRT/MRC}.
\end{aligned}
\right.
\end{eqnarray}

Note that the desired signal $\sqrt {p_{\rm d}} \rho_k {x_{{\rm d},k}}$ in Eq. (\ref{Eqn:yk}) grows with the increase of $M$ while the variance of the downlink SI and the noise is bounded.

\section{Self-Interference Suppression}

\subsection{SI Suppression with a Large-Scale Shared-Antenna Array}
In the traditional large-scale antenna array model, the interferences are asymptotically orthogonal to the subspace spanned by the desired signals when $M$ approaches infinity. Therefore, the desired signal can be recovered by projecting the received signal in a desired signal subspace with linear processing methods such as ZF or MRT/MRC \cite{Ngo1, Ngo2}.

However, in a shared-antenna full-duplex massive MU-MIMO system, the interferences are different from those in the half-duplex or the separate-antenna full-duplex MIMO system. This is because the receivers are very close to the transmitters and the antenna array scale is large, the SI is much stronger than the other interferences such as the inter-user interference in the TDD half-duplex system.  Moreover,  the uplink channel is completely dependent with the downlink channel because the transmitter and receiver are using the same antenna array, which is different from the separate-antenna full-duplex system such as \cite{Ngo2}. Even so, the further study shows that the asymptotic orthogonality still holds true in this new model, namely, the SI can be suppressed by the projection.

\begin{lemma}\label{LEM:lem1}
Let $\bm B$ denote an $M\times M$ complex matrix which satisfies $tr (\bm B\bm B^H)=O(M^2)$; Let $\bm x\triangleq[x_1, ..., x_M]^T$ and $\bm y\triangleq[y_1, ..., y_M]^T$ be $M \times 1$ complex Gaussian random vectors of i.i.d entries which have zero means and unit variances. The entries of $\bm x$, $\bm y$ are mutually independent, and they are all independent of $\bm B$. With $M$ approaching infinity, we have
{\setlength\abovedisplayskip{3pt}
\setlength\belowdisplayskip{5pt}
\begin{align}
&{1 \over M^{3/2}} {\bm x^H\bm B\bm x^*}\xrightarrow{a.s.}0\label{Eqn:xBx*}, \\
&{1 \over M^{3/2}} {\bm x^H\bm B\bm y}\;\;\xrightarrow{a.s.}0\label{Eqn:xBy}.
\end{align}}
\end{lemma}

\begin{IEEEproof}
Assume $\bm x_{\rm R}$ and $\bm x_{\rm I}$ are the real part and imaginary part of $\bm x$. They are mutually independent and have the same distribution. Consider ${1\over M^{3/2}}\bm x^H\bm B\bm x^*$, we get
\begin{align}\label{Eqn:xrxi}
{1\over M^{3/2}}\bm x^H\bm B\bm x^*
&={1\over M^{3/2}}(\bm x{_{\rm R}^H}\bm B \bm x{_{\rm R}} - \bm x{_{\rm I}^H}\bm B \bm x{_{\rm I}}\nonumber\\
&\quad- j\bm x{_{\rm R}^H}\bm B \bm x{_{\rm I}} - j\bm x{_{\rm I}^H}\bm B \bm x{_{\rm R}}).
\end{align}
Like the proof of Lemma 13 in \cite{Hoydis}, we have
\begin{align}\label{Eqn:xr}
&\mathbb{E}\bigg\{\left|{1\over M^{3/2}}\bm x{_{\rm R}^H}\bm B \bm x{_{\rm R}}- {1\over M^{3/2}}tr\bm B\right|^4\bigg \}\nonumber\\
&\leq {C_4\over M^6}(tr\bm B\bm B^H)^2(v{_4^2}+v_8)=O({1/ M^2}),
\end{align}
where $C_4$ is a constant, $v_4$ and $v_8$ are the $4$th and $8$th moments for the entries of $\bm x_R$. Therefore, the right-hand-sdie (RHS) of (\ref{Eqn:xr}) is summable. By Lemma 3 in \cite{Hoydis}, it follows that
${1\over M^{3/2}}\bm x{_{\rm R}^H}\bm B \bm x{_{\rm R}}- {1\over M^{3/2}}tr\bm B\xrightarrow{a.s.}0$.
The same result can be obtained for $\bm x_{\rm I}$. Like the proof of Theorem 3.7 in \cite{Couillet}, we have
${1\over M^{3/2}}\bm x{_{\rm R}^H}\bm B \bm x{_{\rm I}}\xrightarrow{a.s.}0$.
Substitute all the results into Eq. (\ref{Eqn:xrxi}), we have (\ref{Eqn:xBx*}). Similarly, we can prove (\ref{Eqn:xBy}).
\end{IEEEproof}
\emph{Remark:} The results hold true for either the deterministic or the random matrix $\bm B$.
\begin{theorem}\label{Thm:thm1}
In the shared-antenna full-duplex massive MU-MIMO system, the direct-path and reflected-path interference of the uplink SI channel are both vanished if $M$ and $K$ satisfy $M \gg K$ and $M$ is large enough.
\end{theorem}

\begin{IEEEproof}
We consider the SI term in Eq. (\ref{Eqn:r})
\begin{align}\label{Eqn:pd}
\sqrt {p_{\rm d}}\bm W{\!^T} \!{\bm G_{\rm s}}\bm A \bm x_{\rm d}\!=\!\sqrt{p_{\rm d}}\bm W{\!^T}\!\bar{\bm G_{\rm s}}{\bm A}{\bm x_{\rm d}}\!+\!\sqrt{p_{\rm d}}\bm W{\!^T}\!\tilde{\bm G_{\rm s}}{\bm A}{\bm x_{\rm d}}.
\end{align}
For the ZF precoding and decoding case, by substituting Eq. (\ref{Eqn:A}) and Eq. (\ref{Eqn:WT}) into Eq. (\ref{Eqn:pd}), we obtain the first term in the RHS of Eq. (\ref{Eqn:pd}), i.e., the direct-path SI
\begin{align}\label{Eqn:WTGSAZF}
\sqrt{p_{\rm d}}\bm W^T\!\bar {\bm G_{\rm s}}{\bm A}{\bm x_{\rm d}}
\!&=\!\alpha_{\rm ZF}\!\sqrt{p_{\rm d}}{({ {\bm G}}^H \!{{\bm G}})^{\!-\!1}{ {\bm G}}^H}\!\bar {\bm G_{\rm s}}{{{\bm G}}^*}({\bm G}^T\!{\bm G}^*)^{\!-\!1}{\bm x_{\rm d}}\nonumber\\
&=\sqrt{p_{\rm d}}\sqrt {M-K\over M\sum_{k=1}^K{\beta}{_k^{-1}}}\bigg({{{\bm G}}^H {{\bm G}}\over M}\bigg)^{-1}\nonumber\\
&\quad\times \bigg({{ {\bm G}}^H\bar {\bm G_{\rm s}}{{{\bm G}}^*}\over M^{3/2}}\bigg)\bigg({{\bm G}^T{\bm G}^*\over M}\bigg)^{-1}{\bm x_{\rm d}}.
\end{align}
All the factors will converge to a non-zero constant or constant matrices except the $({{{\bm G}}^H\bar {\bm G_{\rm s}}{{{\bm G}}^*}\over M^{3/2}})$ when $M$ is large enough.
Rewrite $\bm G$ as the column form, then we have
\begin{align}
{{ {\bm G}}^H\bar {\bm G_{\rm s}}{{ {\bm G}}^*}}={[{\bm g}{_1^*},{\bm g}{_2^*},...,{\bm g}{_K^*}]^T{\bar {\bm G}_{\rm s}}[{\bm g}{_1^*},{\bm g}{_2^*},...,{\bm g}{_K^*}]},
\end{align}
where ${\bm g}^*_i$ is the $i$th column of matrix ${\bm G^*}$, and $i=1,...,K$.
Note that $\bar {\bm G_{\rm s}}$ satisfies the condition of Lemma \ref{LEM:lem1} by assuming all entries to be $c_{max}$, namely, the worst case of direct-path SI. By using Lemma \ref{LEM:lem1},  we get
\begin{align}
{{\bm g{_i^H}}\bar {\bm G_{\rm s}}{\bm g{_j^*}}\over M^{3/2}}\xrightarrow{a.s.} 0, \quad i,j=1,...,K.
\end{align}
Therefore, we have
\begin{align}\label{Eqn:InDis}
{{ {\bm G}}^H\bar {\bm G_{\rm s}}{{ {\bm G}}^*}\over M^{3/2}}\xrightarrow{a.s.}\bm 0_{K\times K}.
\end{align}
Likewise, we can prove that the reflected-path SI term in Eq. (\ref{Eqn:pd}) is vanished. For the MRT/MRC processing case, we can multiply both sides of Eq. (\ref{Eqn:r}) by $1/M$, then we obtain the result in a similar way.
Thereby, with ZF or MRT/MRC linear processing, the SI will be vanished like the interference and noise in the TDD massive MIMO system.
\end{IEEEproof}

\subsection{The Uplink Decoded Signal}
With the large-scale antenna processing methods, the user data in the uplink received signal can be acquired when $M$ is large enough.
\begin{proposition}
In the shared-antenna full-duplex massive MU-MIMO system, if $M$ and $K$ satisfy $M \gg K$ and $M$ is large enough, the uplink decoded signal for the $k$th user is given by
\begin{align}
r_k\xrightarrow{a.s.}\sqrt{p_{\rm u}}x_{{\rm u},k},
\end{align}
for the ZF precoding and decoding, and
\begin{align}
{1\over{M\beta_k}}{r_k}\xrightarrow{a.s.}\sqrt{p_{\rm u}}x_{{\rm u},k},
\end{align}
for the MRT precoding and MRC decoding.
\end{proposition}
\begin{IEEEproof}
 From Eq. (\ref{eqn:rk}) and Theorem \ref{Thm:thm1}, by using the asymptotic orthogonality property in \cite{Rusek}, i.e., ${1\over M}\bm g{_k^T}\bm g{_k^*}\xrightarrow{a.s.}\beta_k$ and ${1\over M}\bm g{_i^T}\bm g{_j^*}\xrightarrow{a.s.}0$, $i\neq j$, we obtain the results.
\end{IEEEproof}

\subsection{The Downlink Decoded Signal}
With the large-scale antenna processing methods, the downlink SI can also be suppressed.
\begin{proposition}
In the shared-antenna full-duplex massive MU-MIMO system, if $M$ and $K$ satisfy $M \gg K$ and $M$ is large enough, the downlink decoded signal for the $k$th user is given as
\begin{equation}
{1\over \rho_k}{y_{k}} \xrightarrow{a.s.} \sqrt {p_{\rm d}} {x_{{\rm d},k}},
\end{equation}
where $\rho_k$ is given by Eq. (\ref{Eqn:rhok}), it is a linearly increasing function of $\sqrt{M}$ when $M$ is large.
\end{proposition}
\begin{IEEEproof}
The SI $\sum_{q=1}^{K}(c'_{kq}+\beta'_{kq}h'_{kq})$ and the noise term ${n_{{\rm d},k}}$ in the RHS of Eq. (\ref{Eqn:yk}) are both \emph{Gaussian} random variables with finite variance, and the number of UEs K is finite as well, if we multiply both side of Eq. (\ref{Eqn:yk}) by $1\over \rho_k$, we get the result.
 \end{IEEEproof}

The result above looks fantastic. An intuitive explanation is, the received power at each user grows with the increase of the array scale $M$, while the downlink SI is bounded, therefore the user's signal-to-SI ratio grows with the increase of $M$, namely, the SI can be suppressed by the large-scale antenna array. 

%
\section{Performance Results}
In the simulation, we neglect the insertion loss of the circulators and let the number of UEs $K=4$. The large-scale fading $\beta_k$, $\beta$, and $\beta'_{pq}$ are set as 0.1, 0.8, and 0.7, respectively.

\begin{figure}
\centering \scalebox{0.50}{\includegraphics[width=7.6in]{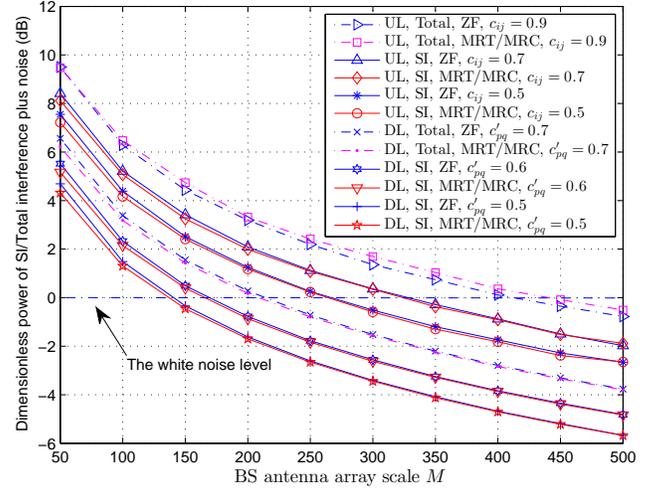}}
\caption{The full-duplex uplink (UL) and downlink (DL) SI or total interference plus noise versus the antenna array scale $M$. The normalized power of uplink and downlink are set as $p_{\rm u}$ = 10 dB and $p_{\rm d}$ = 13 dB, respectively.}\label{fig2}
\end{figure}

Fig. \ref{fig2} simulates the uplink and downlink SI versus the antenna array scale $M$. The vertical axis represents the mean square (the power) of the SI and the total interference plus noise which are from Eq. (\ref{eqn:rk}) and Eq. (\ref{Eqn:yk}).
We can see that the SI and the total interference plus noise level get smaller with the increase of the array scale $M$, the SI level becomes lower with the decrease of the coefficient $c_{ij}$ or $c'_{pq}$.
As shown in Fig. \ref{fig2}, the power of the SI converges to zero at a rate of $1/M$. It will decrease to the level $0$ dB, i.e., the normalized white noise level, at $M\approx260$ for the uplink with $c_{ij}=0.5$ and $M\approx170$ for the downlink with $c'_{pq}=0.6$. The total interference plus noise will decrease to 0 dB at $M\approx430$ for the uplink ($c_{ij}=0.9$) and $M\approx220$ for the downlink ($c'_{pq}=0.7$). Thereby, the SI can be suppressed effectively by the ZF and MRT/MRC methods when $M$ is large enough. We also can see, the ``large enough" $M$ is sensitive to the direct-path SI coefficients $c_{ij}$ and $c'_{pq}$, thus the physical isolation or the duplex mode scheduling for the SI pre-suppression is necessary so that we can get a smaller but ``large enough" $M$.


%

\section{Conclusion}
We proposed a shared-antenna full-duplex massive MU-MIMO model and discussed the uplink and the downlink self-interference suppression. This new model takes advantage of both the high spectral efficiency of the full-duplex system and the channel reciprocity of the TDD massive MIMO system. During the discussion of the SI suppression, we focus on the ZF and the MRT/MRC linear processing methods. A rigorous proof is given to demonstrate that the SI will be suppressed in the completely
dependent uplink and downlink channels when the antenna array scale grows large enough.

\end{document}